\documentclass[conference]{IEEEtran}
\IEEEoverridecommandlockouts
\usepackage{cite}
\usepackage{amsmath,amssymb,amsfonts}
\usepackage{algorithmic}
\usepackage{graphicx}
\usepackage{textcomp}
\usepackage{xcolor}
\usepackage{float}
\def\BibTeX{{\rm B\kern-.05em{\sc i\kern-.025em b}\kern-.08em
    T\kern-.1667em\lower.7ex\hbox{E}\kern-.125emX}}
\usepackage{tikz}
\usepackage{textcomp}
\usepackage{hyperref}
\usepackage{lipsum}

\newcommand\copyrighttext{%
  \footnotesize \textcopyright 2018 IEEE. Personal use of this material is permitted.
  Permission from IEEE must be obtained for all other uses, in any current or future
  media, including reprinting/republishing this material for advertising or promotional
  purposes, creating new collective works, for resale or redistribution to servers or
  lists, or reuse of any copyrighted component of this work in other works. \\
  Accepted to be Published in: Proceedings of the 2018 IEEE Siberian Symposium on Data Science and Engineering (SSDSE),
October 30-31, 2018, Novosibirsk State University, Novosibirsk, Russia}
\newcommand\copyrightnotice{%
\begin{tikzpicture}[remember picture,overlay]
\node[anchor=south,yshift=10pt] at (current page.south) {\fbox{\parbox{\dimexpr\textwidth-\fboxsep-\fboxrule\relax}{\copyrighttext}}};
\end{tikzpicture}%
}
\begin{document}

\title{Reducing over-smoothness in speech synthesis using Generative Adversarial Networks\\
}

\author{\IEEEauthorblockN{ Leyuan Sheng}
\IEEEauthorblockA{\textit{Department of Mathematics and Mechanics} \\
\textit{Novosibirsk State University}\\
Novosibirsk, Russia \\
l.sheng@g.nsu.ru}
\and
\IEEEauthorblockN{ Evgeniy N. Pavlovskiy}
\IEEEauthorblockA{\textit{Stream Data Analytics and Machine Learning laboratory} \\
\textit{Novosibirsk State University}\\
Novosibirsk, Russia \\
pavlovskiy@post.nsu.ru}}

\maketitle
\copyrightnotice
\begin{abstract}
Speech synthesis is widely used in many practical applications. In recent years, speech synthesis technology has developed rapidly. However, one of the reasons why synthetic speech is unnatural is that it often has over-smoothness. In order to improve the naturalness of synthetic speech, we first extract the mel-spectrogram of speech and convert it into a real image, then take the over-smooth mel-spectrogram image as input, and use image-to-image translation Generative Adversarial Networks(GANs) framework to generate a more realistic mel-spectrogram. Finally, the results show that this method greatly reduces the over-smoothness of synthesized speech and is more close to the mel-spectrogram of real speech.
\end{abstract}

\textit{\textbf{Keywords:}} \textbf{\textit{Speech synthesis, over-smoothness, mel-spectrogram, GANs} }

\section{introduction}

Speech synthesis is a technique that produces artificial speech. It is the core of text-to-speech(TTS) and has a wide range of applications: voice assistants, smart stereos, car voice navigation, etc. 

Over time, different technologies have been applied to the field of speech synthesis. Concatenative TTS with HMM-based unit selection, the process of splicing together basic units from a large number of prerecorded voices was the most advanced technology for many years. Parameter TTS was used to generate speech parameters (including fundamental frequency, formant frequency, etc.) at every moment according to a statistical model, and then convert these parameters into waveforms, thus providing high intelligible and fluent speech\cite{c5,c6}.  However, comparing with human speech, the audio produced by these systems usually sounds muffled and unnatural.

In recent years, TTS systems based on deep learning have been extensively studied, and some scientific works have produced surprisingly clear speech. Deep neural network based TTS system includes WaveNet \cite{c7}, Char2Wav \cite{c8}, DeepVoice1\&2 \cite{c9, c10}, and Tacotron1\&2 \cite{c11,c12}. However, there is still a certain gap with real speech, where in the important cause is  over-smoothness in synthesized speech produced form TTS. 

Recently, we have seen emerging solutions for excessive smoothing problems.
In order to close the gap,  Takuhiro Kaneko et al. have incorporated a Generative Adversarial Networks(GANs) as postfilter for short-term Fourier transform(STFT) spectrograms \cite{c13}. However, they used a $F_0$ information and linguistic properties as input features to predict STFT spectra, a very high-dimensional data, not easily trainable by GAN. Ori Ernst  et al. proposed a speech enhancement method in\cite{c133} and D. Michelsanti  et al. proposed a speech dereverberation method in \cite{c14}, they are also using GAN treated spectrogram as an image, but in order to adapt the input shape of GAN, they removed the high frequency bin. Even for testing, they needed to zero-padd the spectrogram of each test sample.

In this paper, our main contribution is to propose a way that converts speech processing into image processing. Based on an end-to-end TTS model Tacotron 2, we use the texts as input to pridict the mel-spectrograms. Then we convert the normalized mel-spectrograms into images, for training and testing to reshape the images into the same size. We then use a GAN framework to generate more realistic mel-spectrograms, in order to overcome the over-smoothness in synthesized speech. 


\begin{figure*}[]
\begin{tabular}{cc}
\begin{minipage}[!tbph]{3.5in}
	\includegraphics[scale=0.9]{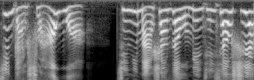}
    \includegraphics[scale=0.9]{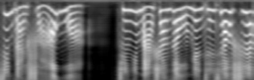}
    \includegraphics[scale=0.9]{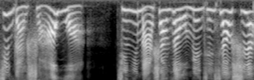}
    \caption{Normalized images of original mel-spectrogram, TTS synthesized mel-spectrogram and GAN-based mel-spectrogram)  }
 	\label{fig:grey}
\end{minipage}
\begin{minipage}[!tbph]{3.5in}
	\includegraphics[scale=0.9]{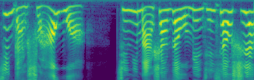}
    \includegraphics[scale=0.9]{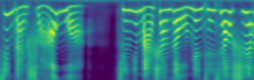}
    \includegraphics[scale=0.9]{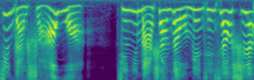}
    \caption{Original mel-spectrogram, TTS synthesized mel-spectrogram and GAN-based mel-spectrogram}
    \label{fig:mels}
\end{minipage}

\end{tabular}
\end{figure*}
\section{Related Works}
\textbf{End-to-End TTS}  Tacotron 1\cite{c11} and Tacotron 2\cite{c12} are end-to-end models which are trained directly from text-audio pairs, greatly reducing dependence on hand-designed internal modules. One advantage of Tacotron 2 over Tacotron 1 is that the speech signal is generated from conditioning the Mel spectrogram which captures all content information, such as pronunciation, prosody, and speaker identity. Tacotron 2 also achieves state-of-the-art sound quality which uses a sequence-to-sequence model optimized for TTS to map a range of functions from sequential letters to encoded audio. These features are an 80-dimensional mel-spectrogram which calculates frames every 12.5 milliseconds, capturing not only the pronunciation of words, and the subtleties of human speech, including volume, speed, and intonation.

\textbf{Generative Adversarial Networks}  Generative adversarial networks(GANs) is a generative model designed by Goodfellow et al. in 2014 \cite{c15}. In the GAN setup, the two differentiable functions represented by the neural network are locked in the game. These two participants, the generator G and the discriminator D, play different roles in this framework. The generator G attempts to generate data from certain probability distribution. The discriminator D, like a judge, determines whether its input is from the generator G or from the real training set.

So far, GAN has been widely used in the field of computer vision and image processing, and has achieved many amazing results.\cite{c16,c17,c18,c19} . Pix2pixHD \cite{c20} is the state-of-the-art for image-to-image translation.

\section{Pix2PixHD framework for Mel-spectrogram}
Mel-spectrogram is a very low-dimensional representation of linear-frequency spectrogram containing spectral envelopes and harmonic information.
The mel scaled acoustic features are the logarithmic sensitivity to the frequency perception of audio signals which are based on the system of human hearing, having an overwhelming advantage in emphasizing audio details \cite{c21}. The Mel-spectrogram features are calculated by dividing a long signal into frames, windowing, and then performing Fourier transform (FFT) on each frame, then stacking the results of each frame along another dimension, and finally passing the mel-scale filter banks like in Tacotron 2. 

The pix2pixHD method is a Conditional Generative Adversarial Networks (CGANs)  framework for image-to-image translation, which improves the pix2pix network by using coarse-to-fine generator G networks and multi-scale discriminator D networks. In our task, the goal is use the generator G reconstruct natural mel-spectrograms $y$ from synthesized mel-spectrograms $s$, while the discriminator D needs to distinguish its input is synthesized or natural. The generator G and discriminator D are trained  through the following a min-max problem:
\begin{equation}
\min\limits_{G}\max\limits_{D_1,D_2,D_3}\sum\limits_{i=1,2,3}\mathcal{L}_{GAN}(G, D_i)
\end{equation}

where the objective function $\mathcal{L}_{GAN}(G, D)$\textcolor{red}{\footnote{we denote $\mathbb{E}_{(\textbf{x},\textbf{y})} \triangleq \mathbb{E}_{(\textbf{x},\textbf{y})\thicksim p_{data}(\textbf{x}, \textbf{y})}$  and $\mathbb{E}_{\textbf{x}} \triangleq \mathbb{E}_{\textbf{x}\thicksim p_{data}(\textbf{x})} $ for simplity}} is given by
\begin{equation}
\mathbb{E}_{(\textbf{x},\textbf{y})}[log D(\textbf{x},\textbf{y})] + \mathbb{E}_{\textbf{x}}[log(1 - D(\textbf{x}, G(\textbf{x})))]
\end{equation}

We use an existing Tacotron 2 model which uses a neural text to mel-spectrogram  conversion model as the baseline to predict the synthesized mel-spectrograms.
We tried to draw a synthesized mel-spectrogram and saved it as an image. However, since the saved image was multi-colored and had four channels, it is difficult to convert the image into a synthesized mel-spectrogram. Therefore, we normalized the synthesized mel-spectrogram first and then saved it as a grayscale image. In fact, the grayscale image we saved can earlier convert into a synthesized mel-spectrogram. The advantage of this is that in addition to making some scaling changes to the image, we can also restore the synthesized mel-spectrogram features, which makes the speech processing and image processing unified.

\section{ Results}
\textbf{Database}:  we used Chinese speech data which contains spoken audio from a little girl, which contains approximately 17 hours of audio with a sample rate of 48 kHz, and consists of 15,000 utterances.

\textbf{Evaluation}: Figure \ref{fig:grey} shows the comparison of normalized images of original mel-spectrogram(upper image), normalized TTS synthesized mel-spectrogram(in the middle) and normalized GAN-based mel-spectrogram(lower image). We fed the TTS synthesized mel-spectrogram as input to Pix2pixHD framework which generated the GAN-based mel-spectrogram. Figure \ref{fig:mels} shows the comparison of the original mel-spectrogram(upper image), TTS synthesized mel-spectrogram(in the middle), and GAN-based mel-spectrogram(lower image). In both comparisons, the  GAN-based mel-spectrogram not only emphasized the harmonic structure, but also reproduced the detailed structure similar to the original  mel-spectrogram than the over-smoothed synthetic  mel-spectrogram.

\section{Conclusions}
In this work we  examined  the process of reducing over-smoothness in speech synthesis by a generative adversarial networks. We achieved this, by adapting the Pix2pixHD framework to transform the speech feature processing into image-to-image translation. The image representation of speech features was directly processed, which was an extention of the previous method to the processing of speech features. We also showed the comparison of speech generated by our proposed method is comparable to original speech.

\end{document}